\documentclass[12pt,a4]{article}

\usepackage[dvips]{graphics}
\usepackage{epsfig}

\begin{document}

\begin{center}
{\Huge  Evolution of Robustness to Noise and Mutation in Gene Expression Dynamics}
\end{center}

\begin{Large}
\begin{center}
Kunihiko Kaneko\footnotemark[1]\ \footnotemark[2]{}
\end{center}
\end{Large}
\begin{center}
\noindent
\footnotemark[1] Department of Pure and Applied Sciences, Univ. of
Tokyo, 3-8-1 Komaba, Meguro-ku, \\
Tokyo 153-8902, Japan\\
\footnotemark[2] Complex Systems Biology Project, ERATO, JST, 3-8-1
Komaba, Meguro-ku, \\
Tokyo 153-8902, Japan \\
\end{center}

%\noindent
%{\bf Corresponding Author:} Kunihiko Kaneko\\
%\hspace{2mm}Department of Pure and Applied Sciences, Univ. of Tokyo,\\
%\hspace{2mm}Komaba, Meguro-ku, Tokyo 153-8902, Japan\\
%\hspace{2mm}Tel/FAX: +81-3-5454-6746\\
%\hspace{2mm}E-mail: kaneko@complex.c.u-tokyo.ac.jp

%\pagebreak
%\noindent

{\Large \bf Abstract}

\noindent
Phenotype of biological systems needs to be robust against mutation in order
to sustain themselves between generations.  On the other hand, phenotype of
an individual also needs to be robust against fluctuations of both internal
and external origins that are encountered during growth and development.  Is
there a relationship between these two types of robustness, one during a
single generation and the other during evolution?  Could stochasticity in
gene expression have any relevance to the evolution of these robustness?
Robustness can be defined by the sharpness of the distribution of
phenotype; the variance of phenotype distribution due to genetic variation
gives a measure of `genetic robustness' while that of isogenic individuals
gives a measure of `developmental robustness'.  Through simulations of a
simple stochastic gene expression network that undergoes mutation and
selection, we show that in order for the network to acquire both types of
robustness, the phenotypic variance induced by mutations must be smaller
than that observed in an isogenic population.  As the latter originates from 
noise in gene expression, this signifies that the genetic robustness evolves 
only when the noise strength in gene expression is larger than some threshold.  
In such a case, the two variances decrease throughout the evolutionary time 
course, indicating increase in robustness.  The results reveal how noise that 
cells encounter during growth and development shapes networks' robustness to 
stochasticity in gene expression, which in turn shapes networks' robustness to 
mutation.  The necessary condition for evolution of robustness as well as 
relationship between genetic and developmental robustness is derived 
quantitatively through the variance of phenotypic fluctuations, which are 
directly measurable experimentally.
                                                                                                        
\pagebreak

\section{Introduction}

Robustness is ability to function against changes in the parameter of a 
system\cite{Barkai-Leibler,Robustness-Alon,Wagner,Wagner2,Evolution}. 
In a biological system, the changes have two distinct origins, 
genetic and epigenetic.  The former concerns with genetic robustness, i.e., 
rigidity of phenotype against mutation, which is necessary to maintain a 
high fitness state.  The latter concerns with fluctuation in number of 
molecules and external environment.

Indeed, phenotype of isogenic individual organisms is not necessarily
identical.  Chemotaxis\cite{Koshland}, enzyme activities, and
protein abundance\cite{Elowitz,Collins,Furusawa,Barkai,noise-review}
differ even among those sharing the same genotype.  Recent studies on
stochastic gene expression elucidated the sources of fluctuations
\cite{Elowitz}.  The question most
often asked is how some biological functions are robust to phenotypic
noise\cite{noise-review,Ueda}, while there may also be
positive roles of fluctuations in cell differentiation, pattern
formation, and adaptation\cite{Kashiwagi,Sawai,IDT,book}.

Noise, in general, can be an obstacle in tuning a
system to the fittest state and maintaining it there.  
Phenotype of an organism is often reproducible even under fluctuating environment
or under molecular fluctuations\cite{Robustness-Alon}.
Therefore, phenotype that is concerned with fitness is expected to
keep some robustness against such stochasticity in gene expression, i.e., 
robustness in `developmental' dynamics to noise.  Phenotype having a
higher fitness is maintained under noise.
How is such ``developmental robustness" achieved through evolution?
In the evolutionary context, on the other hand,
another type of robustness, robustness to mutation need to be considered.
When genetic changes occur, gene expression dynamics are perturbed
so that phenotype with a high fitness  may no longer be maintained.
The ``genetic robustness"
concerns with the stability of a high-fitness state against mutation.

Whether these two types of robustness emerge under natural selection have long been
debated in the context of developmental
dynamics and evolution theory\cite{Wagner,Evolution,Ancel-Fontana,Kirschner},
since the proposition of stabilization selection
by Schmalhausen\cite{Schmalhausen}  and canalization by Waddington\cite{Waddington,GPWagner,Bergman}.
Are developmental robustness to noise and genetic robustness to mutation related?
Is phenotypic noise relevant to attain robustness to mutation?
In the present paper, we answer these questions quantitatively
with the help of dynamical network model of gene expression.

Under the presence of noise in gene expression, phenotype as well as fitness, of isogenic
organisms is distributed, usually following a bell-shaped probability function.
When the phenotype is less robust to noise,
this distribution is broader.  Hence, the variance of this distribution, 
i.e., variance of isogenic phenotypic fluctuation denoted as $V_{ip}$,
gives an index for  robustness to noise in developmental
dynamics.  On the other hand,
robustness to mutation is measured from the fitness distribution over
individuals with different genotypes.  An index for it is given by
variance of phenotypic fluctuation arising from diversity of genotypes
in a population\cite{Fisher,Fisher2,Futuyma}, denoted here as
$V_g$.   This variance $V_g$ increases as the fraction of low-fitness mutants increases.

%Through numerical simulations of the gene network model and
%also from an analysis on the evolution dynamics of fitness  distribution, 
Here we show that evolution to increase both types of robustness
is possible only when  the inequality $V_{ip} \geq V_g$ is satisfied.
Since the isogenic phenotypic fluctuation $V_{ip}$ increases with noise, this means that
evolution of robustness is possible only when the amplitude of phenotypic noise is larger
than some critical value as derived by $V_{ip}\geq V_g$,
implying a positive role of noise to evolution.
We demonstrate that both the two variances $V_{ip}$ and  $V_g$
decrease in the course of evolution, while keeping the proportionality between the two.
This proportionality is consistent with an observation in a bacterial evolution experiment
%, concerning with  evolutionary speed and the variance of the phenotypic fluctuation among 
%clonal individuals
\cite{Sato,book,Kaneko}.

We explain the origin of the critical noise
strength, by noting that smooth dynamical behavior free from a rugged
potential landscape evolves as a result of phenotypic noise.
When the noise amplitude is smaller than the threshold,
we observe that low-fitness mutants are accumulated, so that
robustness to mutation is not achieved.
Generality and relevance of our results to biological evolution are
briefly discussed.

\section{Theoretical Framework on Genetic-Phenotypic Relationship}

In natural population, both the phenotype and genotype differ among 
individuals.  Let us consider population
distribution $P(x,a)$ where $x$ is a variable characterizing a phenotype 
and $a$ is that for the corresponding genotype\cite{Kaneko}.  Here the phenotype $x$ is
responsible for the fitness of an individual, and the selection depending on  $x$ is 
considered as an evolutionary process.
Since the phenotype differs even among isogenic individuals, the distribution
$P(x;a=a_0)$ for a fixed genotype $a_0$ has some variance.  This isogenic
phenotypic variance $V_{ip}$, defined as the variance
over clones, is written as $V_{ip}(a)=\int (x-\overline{x(a)})^2 P(x,a)dx$, where
$\overline{x(a)}$ is the average phenotype of a clonal population
sharing the genotype $a$, namely $\overline{x(a)}=\int P(x,a)x dx$.
This variation of phenotype is a result of noise through the developmental process
to shape the phenotype.  If this variance is smaller, the phenotype is less influenced
by noise, and thus $V_{ip}$ works as a measure of robustness of the phenotype against noise.

On the other hand, the standard evolutionary genetics \cite{Fisher,Fisher2,Futuyma}
mainly studies the phenotypic variance due to genetic variation. It
measures phenotypic variability due to diversity in genotypes in a population.
This phenotypic variance by genetic variance, which is termed $V_g$ here,
is then defined as the variance of the average $\overline{x(a)}$,
over genetically heterogeneous individuals.  It is given by
$V_{g}=\int (\overline{x(a)}-<\overline{x}>)^2 P(a)da$,
where $P(a)$ is the distribution of the genotype $a$
and $<\overline{x}>$ is the average of $\overline{x(a)}$ over genotypes $a$.
While $V_{ip}$ is defined as variance over clones, i.e., individuals with the same genotype,
$V_g$ comes from those with different genotypes.
As $V_g$ is smaller, the phenotypic change by genetic variation is smaller.
Hence $V_g$ gives a measure of robustness of the phenotype against mutation. 

%\footnote{ Strictly speaking, $V_g$ is defined as variance of {\em average}  phenotype 
%exhibited by individuals of diverse genotype.}.

%On the other hand, the non-genetic part of a phenotypic fluctuation as a result of noise in developmental
%process is termed as the isogenic phenotypic variance, $V_{ip}$ here.  

%Here the phenotype $x$ can be regarded as fitness of an individual,
%and the selection for $x$ is considered as an evolutionary process.
%Now, $V_{ip}$ is the variance of $x$ and this can be written as
%$V_{ip}(a)=\int (x-\overline{x(a)})^2 P(x,a)dx$, where
%$\overline{x(a)}$ is the average phenotype of a clonal population
%sharing the genotype $a$, namely $\overline{x(a)}=\int P(x,a)x dx$.
%$V_g$ is then defined as the variance of the average $\overline{x(a)}$,
%over genetically heterogeneous individuals, given by
%$V_{g}=\int (\overline{x}-<\overline{x}>)^2 P(\overline{x})d\overline{x}$,
%where $P(\overline{x})$ is a distribution of average phenotype,
%and $<\overline{x}>$ as the average of $\overline{x(a)}$ over genotypes.
%As a typical value of $V_{ip}$, we take its value at $a=a_0$, where $P(x,a)$ is maximal.
%( see also Supporting text).

%\footnotetext{ As a typical value of $V_{ip}$, we take its value at the genotype $a=a_0$, where $P(x,a)$ is maximal.}, 

We have previously derived  an inequality $V_{ip}>V_g$ between the two variances,
by assuming evolutionary stability of  the population
distribution $P(x,a)$, that is preservation of single-peakedness
through  the course of evolution \cite{Kaneko} (see Supporting text).  
Indeed the single-peaked distribution collapses as $V_{ip}$ approaches  $V_g$, where
the distribution is extended to very low values of $x$ (fitness).  In other words,
error catastrophe occurs at  $V_g \approx V_{ip}$;
(Here error catastrophe\footnotemark
means accumulation of low-fitness mutants in the population after generations.)
For each course of evolution under a fixed mutation rate,
the proportionality between $V_g$ and $V_{ip}$ is derived,
since the genetic variance increases roughly proportionally to the mutation rate\cite{Kaneko}.
%To sum up, by assuming two variable distribution and evolutionary stability,
%the following are derived (see Supporting text):

\footnotetext{The term means accumulation of low-fitness mutants in the population after generations, and
is used here by extending its original meaning by Eigen\cite{Eigen}.}

%(i) $V_{ip} \geq V_g$

%(ii) error catastrophe at  $V_g \approx V_{ip}$;

%(iii) proportionality between $V_g$ and $V_{ip}$ through a given course of evolution

Note, however, that the derivation of these relationships
($V_{ip} \geq V_g$, error catastrophe at $V_g \approx V_{ip}$, and proportionality
 between $V_g$ and $V_{ip}$ for a given course of evolution) is based on
the existence of two-variable distribution function $P(x=phenotype,a=gene)$, and
the postulate that single-peaked distribution is maintained
throughout evolution, which is not trivial.
Hence the above relationships need to be examined by some
models for evolution.
In addition,
{\sl why does the population distribution extend to low-fitness values
when the phenotypic fluctuation $V_{ip}$ is smaller than $V_g$}?
Or, put it another way, why do systems with small phenotypic noise
run into "error catastrophe"?
In fact, the emergence of error catastrophe as a result of decreasing
isogenic phenotypic fluctuation below $V_g$ may look rather counter-intuitive,
since in general one expects fluctuation to perturb a system from the fittest state.  The necessity of fluctuation
for evolution to increase robustness to noise and to mutation needs theoretical
examination.

\section{Model}

To study the proposed relationships, we need to consider seriously how the phenotype is shaped through
complex ``developmental process".  In the present paper, we use the term `development',
in a broad sense, including a process in uni-cellular organisms to reach 
cell division.  It is a dynamical process to shape a phenotype at a 'matured' state (where fitness is defined)
from a given initial state.  In general,  this dynamic process is complex so that the process
may not reach the identical phenotype due to the noise through this developmental process.
This leads to the isogenic variance of the phenotype $V_{ip}$.  On the other hand,
the equation governing the developmental process is varied as a result of mutation.  
The phenotype variance over a population with distributed genotypes gives $V_g$.

We consider a simple model to satisfy the requirement on 'development' above. It
consists of a complex dynamic process to reach a target phenotype under a noise which may
alter the final phenotypic state.  We do not choose a biologically realistic model that describes
a specific developmental process, but instead take a model as simple as possible, to satisfy
a minimal requirement for our study.  Here we take a simplified model, borrowed
from a gene regulatory network, where expression of a gene activates or inhibits expression of other genes under noise.
These interactions between genes are determined by the network. The expression profile
changes in time, and eventually reaches a stationary
pattern.  This gene expression pattern determines fitness.
%The network is determined genetically, and
Selection occurs after introduction of mutation at each generation in the
gene network.  Among the mutated networks, we select a network with a
higher fitness value.  Since there is a noise term
in the gene expression dynamics, fitness fluctuates even among the individuals with an
identical gene network, which leads to the isogenic fluctuation $V_{ip}$.
On the other hand, the expression pattern varies by mutation in the network,
and gives rise to variation in the average fitness, resulting in $V_g$.

This simplified gene expression follows a typical switch-like dynamics with a sigmoid input-output behavior
\cite{gene-net,Mjolsness,Sole,Newman,threshold}
widely applied in models of signal transduction\cite{signal} and neural networks\cite{neural}
(For a related evolution model with discrete states, see e.g., \cite{Bergman}).
The dynamics of a gene expression level $x_i$ is described by

\begin{equation}
dx_i/dt =\tanh[\beta \sum_{j >k}^{M} J_{ij}x_j]-x_i +\sigma \eta(t),
\end{equation}

\noindent
where $J_{ij}=-1,1,0$, and $\eta(t)$ is Gaussian white noise given by
$<\eta(t)\eta(t')>= \delta(t-t')$.   $M$ is the total number of genes, and $k$ is
the number of output genes that are responsible for fitness to be determined.
The value of $\sigma$ represents noise strength
that determines stochasticity in gene expression\footnote{For simplicity we mainly
consider the case that the noise amplitude is independent of $x_i$, while inclusion of
such $x$-dependence of noise amplitude does not alter the conclusion to be discussed.}.
By following a sigmoid function $tanh$,  $x_i$ has
a tendency to approach either 1 or -1, which is regarded as ``on" or ``off" of
gene expression.  Even though $x$ is defined over $[-\infty,\infty]$, it is attracted
to the range [-1,1] (or slightly above or below the range due to the noise term).
We consider a developmental process leading to a matured phenotype
from a fixed initial state, which
is given by (-1,-1,...,-1); i.e., all genes are off,
unless noted otherwise \footnotemark .
\footnotetext{This specific choice of initial condition is not important.}

Let us define a fitness function so that gene expression levels ($x_i$)
for genes $i=1,2,\cdots,k(<M)$ would reach an ``on'' state, i.e., $x_i>0$. The fitness is maximum
if all $k$ genes are on after a transient time span $T_{ini}$, and minimum if
all are off.  To be specific, we define the fitness function by

\begin{equation}
F= \sum_{j=1}^k ([S(x_j)]_{temp}-1)=\frac{1}{T_f-T_{ini}}\sum_{j=1}^k \int_{T_{ini}}^{T_f} (S(x_j)-1) dt
\end{equation}

\noindent
where $S(x)=1$ for $x>0$, and 0 otherwise,
$[...]_{temp}$  is time  average between  $t=T_{ini}$ and  $t=T_f$\footnote{The time average
here is not important, because the gene expressions $x_i$ are fixed after some time, 
in most cases.  Adoption  of the value $(S(x_j)-1)$ after initial time $T_{ini}$ leads to the 
same result.}.
The fitness function takes the maximum value $F=0$ when the selected pattern of 
gene expression ($x_i$; $i=1,2,\cdots ,k$) is always ``on" 
%for the given time interval 
and takes the minimum ($F=-k$) when all $k$ genes are always off.
Note that fitness is calculated only after 
time $T_{ini}$, which is chosen sufficiently large so that 
the temporal average can be computed
after the gene expression dynamics has fallen on an attractor.
This initial time can be considered as the time required for
developmental dynamics.

As the model contains a noise term, fitness fluctuates at
each run, which leads to the distribution in $F$, even for the same network.
Hence we obtain the distribution $p(F;g)$, for a given network ``g", whose
variance gives isogenic phenotypic fluctuation.
At each generation, we compute the fitness $F$ over $L$ runs, to obtain
the average fitness value $\overline{F}$ of a given network.  

Now we consider the evolutionary process of the network.  Since the network
is governed by $J_{ij}$ which determines the `rule' of the dynamics,
it is natural to treat  $J_{ij}$ as a measure of genotype.
Individuals with different genotype have a different set of $J_{ij}$. 
At each generation there are $N$ individuals with different sets of $J_{ij}$. 
For each individual network, we compute the average fitness $\overline{F}$.
Then we select the  top $N_s (<N)$ networks that have higher fitness values.
(The value $N/N_s$ corresponds to the selective pressure.  As it is larger, the evolution
speed increases. However, specific choice of this value itself is not important to
the result to be discussed).

At each generation,  mutation changes the network, i.e., changes
$J_{ij}$ at a given mutation rate $\mu$. 
We rewire the network at a given rate so that 
changes in $J_{ij}$ produce $N$ new networks.
(In most simulations, only a single path, i.e., a single  pair of $i,j$
is changed. The mutation rate can be lowered by changing a path only for some probability
\footnote{Although the mutation rate is important to the evolution speed and alters the error catastrophe point to be discussed,
the conclusion to be discussed is not altered by specific choice of $\mu$.}.
Here we make $N/N_s$ mutants from each of the top $N_s$ networks, 
so that there will be $N$ networks again for the next generation.
From this population of networks we repeat the process of the developmental dynamics,
fitness calculation, selection, and mutation \footnote{
Instead of this simple genetic algorithm, we can also assume that 
the number of offspring increases with the fitness.  This choice
does not alter the conclusion to be presented.}.

Simulations start from a population of random networks with a given fraction of
paths (for example, 50 \% of $J_{ij}$ are nonzero).  At each
generation, the $N$ individuals have slightly different networks
$J_{ij}$, so that the values of $\overline{F}$ are different.  We
denote the fitness distribution over individuals with different
genotype as $P(\overline{F})$.  On the other hand, the fitness
distribution for an identical network (``g") is computed to obtain
$p(F;g)$.

{\sl Remark}: Developmental dynamics and selection process in our model are
too simple.  Still, this model is relevant to examine general statement on phenotypic fluctuations,
as the model at least captures complex dynamics giving rise to a phenotype, 
stochasticity in dynamics, mutation,  and selection according to a given phenotype. 
Indeed, real gene expression dynamics depend on environmental conditions, and the
fitness is defined as expression patterns to adapt each environmental condition.
We have also carried out some simulations by imposing such fitness but the
results to be discussed (with regards to $V_g$ and $V_{ip}$) are invariant.

\section{Results}
 
Let us first see how the evolutionary process changes as a function of
the noise strength $\sigma$.  After generations, the peak position in $P(\overline{F})$ increases, so that the top
of $\overline{F}$ in the population approaches the highest value $0$. Indeed, in all cases, 
the top group quickly evolves to the highest fitness state $\overline{F}=0$ (see Fig.1a)\footnote{Even for
$\sigma =0.2$, the highest fittest value approaches 0 after a few hundred more generations.}.  The time 
necessary for the system to reach this state becomes shorter
as the phenotypic noise decreases (see Fig.2).
On the other hand, the time evolution of the distribution function $P(\overline{F})$ depends
drastically on the noise strength $\sigma$.  When $\sigma$ is small, 
the distribution is broad and the existing individual with 
the lowest $\overline{F}$ remains at the low fitness state, while
for large $\sigma$, even the individuals with the lowest fitness approach $\overline{F}=0$ (see Fig.1b and Fig.3).
There is a threshold noise $\sigma_c(\approx 0.02)$, below which the distribution $P(\overline{F})$ is
broadened, as is discernible in the data of the variance of the distribution, $V_g$ in Fig.2.
Here, the top individuals reach the highest fitness, leaving
others at the very low fitness state.  As a result,  the average fitness over
all individuals, $<\overline{F}>=\int  \overline{F} P(\overline{F})d\overline{F}$ is low.
$<\overline{F}>$ and the lowest fitness over individuals $\overline{F}_{min}$,
after a sufficiently large number of generations, are plotted against $\sigma$ in Fig.2.  
The abrupt decrease in fitness suggests threshold noise $\sigma_c$, below which
low-fitness mutants always remain in the distribution.   
For $\sigma>\sigma_c$, the distribution $P(\overline{F})$ takes a 
sharp peak at $\overline{F} \sim 0$, where the variance is rather small\footnotemark .
Distribution below and above $\sigma_c$ are displayed in Fig.3.

\footnotetext{This type of transition is also observed by increasing the mutation rate,
while fixing the noise strength at $\sigma>\sigma_c$.}

Let us study the relationship between $V_g$ and $V_{ip}$.  Here $V_{ip}$
is defined as variance from the distribution $p(F;genotype)$, i.e., over individuals with the
same genotype.  As the distribution $p$ depends on each
individual with different genotype, the variance changes accordingly.
Naturally, the top individual has a smaller variance, and the individual
with lower fitness has a larger variance.  As a measure of $V_{ip}$,
we used either the average of the variance over all individuals or
the variance of phenotype from a gene network that is located
closest to the peak in the distribution $P(\overline{F})$.  
Both estimates of
$V_{ip}$ do not differ much, and the following conclusion is drawn in
both cases.
$V_g$, on the other hand, is simply the variance of the distribution $P(\overline{F})$, 
i.e., over individuals having different genotypes present. 

The relationship between $V_g$ and $V_{ip}$ thus evaluated is plotted
in Fig.4.  We find that both the variances decrease through the
evolutionary time course when $\sigma > \sigma_c$, where we note:

(i) $V_{ip}>V_g$ for $\sigma > \sigma_c$.

(ii) $V_g \propto V_{ip}$ during the evolutionary time course under a fixed noise strength
$\sigma(>\sigma_c)$ and a fixed mutation rate.  As $\sigma$ is lowered toward $\sigma_c$, $V_g$ increases so that
it approaches $V_{ip}$.   

(iii) $V_g \approx V_{ip}$ at $\sigma \approx \sigma_c$, where error catastrophe occurs.

In other words, the fittest networks maintaining
a sharp distribution around the peak dominate only when $V_{ip}>V_g$
is satisfied. As $\sigma$ is decreased to $\sigma_c$, $V_{ip}$
approaches $V_g$, error catastrophe occurs and a considerable
fraction of low-fitness mutants accumulates.  Hence, the relationships proposed theoretically 
assuming  evolutionary stability of a two-variable distribution function
$P(x=phenotype,a=genotype)$ is confirmed.
Here, without introducing phenomenological assumptions, the three relationships
are observed in a general stochastic gene-network model.

Why didn't the system maintain the highest fitness state under small
phenotypic noise $\sigma<\sigma_c$?  To study the difference in
dynamics evolved for different values of $\sigma$, we choose the top
individual (network) that has evolved at $\sigma=\sigma_0$, and place
it under a different noise strength $\sigma=\sigma '$.
In Fig.5, we have plotted the fraction of runs giving rise to $F=0$
under such circumstance. As shown, the successful fraction decreases
when $\sigma'$ goes beyond $\sigma_0$.  In other words, the network
evolved under a given noise strength successfully reaches the target gene expression
up to that critical noise level, while it begins to fail doing so
at a higher noise strength.
Accordingly, the distribution $p(F;gene)$ extends to lower values in
fitness, when a network evolved under small phenotypic noise is
developed under a higher noise level.  
On the other hand, the network evolved under high level noise maintains a high
fitness value, even when the noise is lowered.

Next we study the basin structure of
attractors in the present system.  Note that an orbit of the network with the highest
fitness, starting from the prescribed initial condition, is
within the basin of attraction for an attractor corresponding to the target
state ( $x_i >0$ for $i=1,\cdots ,k$).  Hence the basin of attraction for
this target attractor is expected to be larger for the dynamics
evolved under higher level noise.
We have simulated the dynamics (1) for the evolved fittest network 
under zero noise, starting from a variety of
initial conditions over the entire phase space,
and measured the distribution of $F$ at each attractor.
The distribution is shown in Fig.6\footnotemark .  
For the network evolved under $\sigma>\sigma_c$, the distribution has a
sharp peak at $F=0$ (and $F=-k$ due to the symmetry), with more than
40\% attraction to each.  On the other hand, for the networks evolved
under $\sigma< \sigma_c$, the peak height at $F\sim 0$ is very small,
i.e., the basin for the attractor with $F=0$ is tiny.  There
exist many small peaks corresponding to attractors with $-k<F<0$,
having similar sizes of basin volumes.  In fact, the basin volumes for
attractors with $-k<F<0$ grow as $\sigma$ is decreased, and are
dominant for $\sigma <\sigma_c$.

\footnotetext{
Due to the symmetry against $x_j=1 \leftrightarrow x_j=-1$ in the model, the
distribution is symmetric around $F=-k/2$ when all initial conditions
are taken.  In fact, by starting from $x_i=1$ for all $i$, the orbit
reaches an attractor $x_j<0$ for $j=1,\cdots ,k$, resulting in $F=-k$.}  

\section{Dynamic Origin of Robust Evolution}

The difference in the basin structure suggested by Fig.6 is
schematically displayed in Fig.7.  For the network evolved under
$\sigma>\sigma_c$ there is a large, smooth attraction to the target
state, while for the dynamics evolved under $\sigma< \sigma_c$, the
phase space is split into small basins.  Let us consider the distance between the
basin boundaries from a ``target orbit" starting from $-1,...,-1$ and
reaching $x_i>0$ (for $1\leq i \leq k$) which is defined by $\Delta$ here.
The distance $\Delta$ remains small for the dynamics evolved under a low
noise strength $\sigma <\sigma_c$, and increases for
those evolved under a higher noise.
It is interesting to note that evolution influences the basin
structure globally over the phase space, although the fitness
condition is applied locally to an orbit starting from a specific
initial condition.

The results in Fig.5 and Fig.6 imply that the gene regulation networks that operate and 
evolve under noisy environment exhibit qualitatively different dynamics compared to 
those subjected to low level noise.  In our model, the fitness
of an individual changes when its gene expression $x_j$ for $j=1,\cdots ,k$ changes its sign.
Recall that the fixed point solution
$x_i=tanh(\sum_j J_{ij}x_j)$ changes its sign when $\sum_j J_{ij}x_j$
in the sigmoid function  changes its sign.
This change may occur during the developmental dynamics by noise, and we call
such points in the phase space  `turning points'.  
When an orbit of eq.(1)
passes over turning points, $x_j$ takes a negative value for
some $j$ (for $1\leq i \leq k$) at the attractor (see Fig.8 for
schematic representation).
Since there are many variables for gene expression and the values of $J_{ij}$ are distributed
over -1 and 1, the term $tanh(\sum_j J_{ij}x_j)$ changes its sign at several points in
the phase space $\{ x_j \}$ generally.
Hence there can be many turning points in the phase
space.  The fittest network with $\overline{F}\approx0$ chooses such
orbits having no turning points within the noise range
from the original orbit.  An orbit from the fittest individual evolved
under low-level noise encounters many turning points when subjected
to noisy environment.
 
The average distance between the turning
points and an orbit that has reached the target gene expression pattern
is estimated by the distance $\Delta$ defined above.  
Recall that the distance $\Delta$ is small for the dynamics
evolved under a low noise strength.
Such dynamics, if perturbed by a higher level of noise, are
easily caught in the turning points, which explains the behavior
shown in Fig.5.

Let us now discuss the relationship between $V_g$ and $V_{ip}$.  Noise
disturbs an orbit so that it may go across the basin boundary (turning
points) with some probability.  We denote the standard deviation of
the location of the orbit due to noise as $\delta_p$, which is
proportional to the noise strength $\sigma$.  Since the distance
between the orbit and the basin boundary is deviated by $\delta_p$,
and the fitness value drops when the orbit crosses the basin boundary,
the variance $V_{ip}$ is estimated to be proportional to
$(\delta_p/\Delta)^2$.

Next, we discuss how the mutation in the network influences the
dynamics.  When the network is altered, i.e., a path is added or
removed as a result of mutation in $J_{ij}$, there exists a variation
of the order of $O(1/\sqrt{N})$ in the threshold function term in
eq.(1).  This leads to a deviation of the location of turning points
(or basin boundary).  We denote this deviation as $\delta_g$, which
increases with the mutation rate.  The variance $V_{g}$ is estimated
to be proportional to $(\delta_g/\Delta)^2$, with the same proportion
coefficient as that between $V_{ip}$ and $(\delta_p/\Delta)^2$.

Under the presence of noise, there is a selective pressure to avoid
the turning points (basin boundaries) that exist within the distance
$\delta_p$ from the ``target" orbit.  This leads to an increase in
$\Delta$.  However, if $\delta_g$ is larger than $\delta_p$, the
memory of this distance between the target and the boundaries will not
be propagated to the next generation, due to large perturbation to the
original network by the mutation.  Hence increase in $\Delta$ (i.e.,
increase in robustness) is expected only if $\delta_p>\delta_g$.
Since $\delta_p$ and $\delta_g$ increase with the noise strength
$\sigma$ and the mutation rate $\mu$ respectively, there exists a
critical noise strength $\sigma_c$ beyond which this inequality is
satisfied.  From the relationship between $\delta_{p,g}$ and
$V_{ip,g}$, the condition for robust evolution is rewritten as
$V_{ip}>V_g$.

When the condition $V_{ip}>V_g$ (i.e., $\sigma>\sigma_c$) is
satisfied, the system increases $\Delta$ during evolution.  We have
computed the temporal evolution of basin distribution.  With
generations, the distribution evolves from the pattern at a low level
noise in Fig.7, to that at large $\sigma$ characterized by enhanced
peak at $F = 0$.  Accordingly $\Delta$ increases with generations.
Recall that $V_{ip} \sim (\delta_p/\Delta)^2$, and $V_{g} \sim
(\delta_g/\Delta)^2$, both variances decrease with generations, while
$V_{ip}/V_g$ is kept constant.
%This concludes the explanation of the $V_g$-$V_{ip}$ relationships,
%although the argument here requires further mathematical elaboration.

\section{Discussion}

We have demonstrated the inequality and proportionality between $V_g$
and $V_{ip}$, through numerical evolution experiment of a gene
network.  As phenotypic noise is decreased and the inequality
$V_{ip}>V_g$ is broken, low-fitness mutants are no longer eliminated.
This is because the mutants fail to reach the target gene expression
pattern, by crossing the boundary of the basin of attraction to the
target.  When the amplitude of the noise is larger, on the other hand,
the networks of the dynamics with a large basin volume for the target
attractor are selected and thus mutants with lower fitness are removed
successively through the selection process.  Hence noise increases
developmental robustness through evolution, together with genetic robustness.

Although we used a specific example to demonstrate the relationship
between $V_{ip}$ and $V_g$ and the error catastrophe, we expect 
this relationship to be generally applicable to systems
satisfying the following conditions:

(i) Fitness is determined through developmental dynamics.

(ii) Developmental dynamics is sufficiently complex so that its orbit,
when deviated by noise, may fail to reach the state with the highest
fitness.

%(iii) There is similarity between phenotype change induced
%by genetic change and that by noise in the developmental dynamics.

(iii) There is effective equivalence between mutation and noise in the
developmental dynamics with regards to phenotype change.

Note that the present system as well as the previous cell
model\cite{Kaneko} satisfies these conditions.  The condition (i) is
straightforward in our model, and the condition (ii) is satisfied
because of the complex dynamics having many turning points in the
phase space.  Noise in developmental dynamics sometimes perturbs an
orbit to cross the basin boundary so as to escape from attraction to
the target gene expression pattern, while a mutation in the network
may also induce such failure, by shifting basin boundaries. Hence the
condition (iii) is satisfied.

When developmental process fails due to phenotypic noise, the fitness
function takes a low value.  Evolution under noise acts to prevent
such failure within the range of noise.  On the other hand, due to the
condition (iii), mutation may also lead to such lethality. When the
effect of mutation goes beyond the range given by the phenotypic
noise, mutants with very low fitness values begin to accumulate.
Hence there appears a threshold level of phenotypic noise below which
low-fitness (or deleterious) mutants accumulate (or threshold mutation
rate beyond which such mutants accumulate).  In this sense, we expect
that for robust evolution, the inequality $V_g<V_{ip}$ must be
satisfied in order for the low-fitness mutants to be eliminated.
Violation of the inequality leads to accumulation of low-fitness (or
deleterious) mutants, a phenomenon known as error
catastrophe\cite{Eigen}.  Only under the presence of noise in the
developmental process, systems acquire robustness through
evolution. In other words, developmental robustness to stochasticity
in gene expression implies genetic robustness to mutation.  
%This result sheds a new light on the role of developmental noise in the
%evolutionary context\cite{Ancel-Fontana,Evolution,Wagner}.
Quantitative analyses of stochasticity in protein abundance during the
laboratory evolution of bacteria are possible \cite{Sato,Alon}.  By
carefully measuring the variation $V_g$ of given phenotype in mutants,
and comparing it with that of isogenic bacteria, $V_{ip}$, one can
examine the validity of our conclusion between $V_g$ and $V_{ip}$.

It is worthwhile to mention that in a class of theoretical models,
fitness landscape is given as an explicit continuous
function of a gene sequence (e.g., energy
function in a spin glass\cite{SG}), where a minute change in sequence does not lead to
a drastic change in fitness.  On the other hand, in a system
satisfying (i) and (ii), a small change in genotype (e.g., a single
change in the network path) may result in a large drop in 
fitness, since fitness is determined after the developmental
dynamics.  Indeed, there may appear mutants with very low fitness
values from an individual with a high fitness value, only by a single
change of a path in the network.  Such
deleterious mutations are also observed in nature\cite{Futuyma}.

It is interesting to note that a larger basin of attraction to a
target attractor (with the highest fitness value) is formed through a mutation and
selection process.  As a result, dynamics over the entire phase space are
simplified to those having only a few attractors, even though the
fitness function is given locally 
without scanning over the entire phase space.  When the time-course is
represented as a motion 
along a potential surface, our results suggest
that the potential landscape becomes smoother and simpler through evolution, and
loses ruggedness after generations.  Indeed, existence of such global
attraction in an actual gene network has recently been
reported in yeast cell-cycle\cite{Ouyang}.

Such smooth landscape was also studied in protein folding\cite{Go,Wolynes}.
Saito et al.\cite{Sasai} observed an evolutionary
process from a rugged to the so-called funnel-like landscape in an interacting spin 
system abstracting protein folding dynamics.  Under a general  framework of statistical
mechanics\cite{Nishimori}, a relationship between the degree of variance in 
coupling coefficients $J_{ij}$ between spins (corresponding
to $V_g$) and the temperature (i.e., phenotypic noise for spin $x_i$,
corresponding to $V_{ip}$) is formulated.  Such relationship may be relevant to understand the
relationships between $V_g$ and $V_{ip}$ in our study.
%the present context.

According to established Fisher's theorem on natural selection,
evolution speed of phenotype is proportional to the phenotypic
variance by genetic variation, $V_g$\cite{Fisher,Fisher2,Futuyma}.  The demonstrated
proportionality between  $V_{ip}$ and $V_g$ then suggests that the
evolution speed is proportional to the isogenic phenotypic fluctuation,
as is also supported by an experiment on
bacterial evolution in a laboratory\cite{Sato} and confirmed by simulations of
a reaction network model of a growing cell\cite{Kaneko}. 

Isogenic phenotypic fluctuation is related to phenotypic plasticity,
which is a degree of phenotype change in a different environment.
Positive roles of phenotypic plasticity in evolution have 
been discussed\cite{Kirschner,Eberhard,Ancel,plasticity}. 
Since susceptibility to the environmental change and the phenotypic fluctuation
are positively correlated according to
the fluctuation-response relationship\cite{book,Kubo}, our present results
on the relationship  between phenotypic fluctuations and evolution 
imply, inevitably, a relationship between  phenotypic plasticity and evolution 
akin to genetic assimilation proposed by Waddington\cite{Waddington}.

I would like to thank Chikara Furusawa, Koichi Fujimoto, Masashi Tachikawa, Shuji Ishihara,
Katsuhiko Sato, Tetsuya Yomo, Philippe Marcq, Uri Alon, and Satoshi Sawai for stimulating discussion.

\pagebreak

Figure Caption:

Fig.1:
Evolutionary time course of the fitness $\overline{F}$.  The highest (a) and
the lowest (b) values of the fitness $\overline{F}$ among all
individuals that have different genotypes (i,e., networks $J_{ij}$) 
at each generation are plotted.  Plotted are for different
values of noise strength, $\sigma=$ 0.01,0.02,0.04,0.06,0.08,0.1, 0.2
with different color.  

Hereafter we mainly present the numerical results for $M=64$ and
$k=8$. At each generation there are $N$ individuals.  $N_s=N/4$
networks with higher values of $\overline{F}$ are selected for the
next generation, from which mutants with a change in a single element
$J_{ij}$ are generated.  For the average of fitness, $L$ runs are
carried out for each.  Unless otherwise mentioned, we choose
$N=L=300$, while the conclusion to be shown does not change as long as
they are sufficiently large.  (We have also carried out the selection
process by $F$ instead of $\overline{F}$, but the conclusion is not
altered if $N$ is chosen to be sufficiently large.) Throughout the
paper, we use $\beta=7$.

Fig.2: 
Average fitness $<\overline{F}>$, lowest average fitness
$\overline{F}_{min}$, evolution speed, and variance of the fitness $V_g$
are plotted against the noise
strength $\sigma$.  $<\overline{F}>$, the average of the average
fitness $\overline{F}$ over all individuals is computed for 100-200
generations (red cross, from the simulation with population of 100
individuals, and purple square from 300 individuals). The minimal
fitness is computed from the time average of the least fit network present
at each generation (green, from 100 population, and light blue from 300). The
evolution speed is plotted, measured as the inverse of the time
required for the top individual to reach the maximal fitness 0.
$V_g$ is computed as the variance
of the distribution $P(\overline{F})$ at 200th generation.

Fig.3:
Distribution $P(\overline{F})$ after 200 generations, for population
of 1000 individuals.  Inset is the magnification for
$-0.2<\overline{F}<0$.  For high $\sigma $ (red, with $\sigma=0.1$),
the distribution is concentrated at $\overline{F}=0$, while for low
$\sigma$ (green, with $\sigma=0.006$), the distribution is extended to
large negative values, even after a large number of generations.

Fig.4:
Relationship between $V_g$ and $V_{ip}$.  $V_g$ is computed from
$P(\overline{F})$ at each generation, and $V_{ip}$ by averaging the
variance of $p(F;gene)$ over all existing individuals. (We also
checked by using the variance for such gene network that gives the
peak fitness value in $P(\overline{F})$, but the overall relationship
is not altered).  Plotted points are over 200 generations.
For $\sigma> \sigma_c\approx .02$, both decrease with generations.

Fig.5:
Dependence of the fraction of the runs that reach the target
expression pattern. Networks that had the top
fitness value under noise $\sigma$ were simulated at a different noise $\sigma'$.  
We first generate a network as a result of evolution over 200 generations under the noise
strength $\sigma$, and select such network $J_{ij}$ that has the top
fitness value.  Then we simulate this network under new noise strength
$\sigma'$ from the initial condition $-1,-1,\cdots ,-1$ over 10000 runs, to
check how many of them reach the target pattern (i.e., $x_i>0$ for
$i=1$ to 8).  Plotted is the fraction of such runs against the noise
strength $\sigma'$.  Different color corresponds to the value of the
original noise strength $\sigma$ used for the evolution of the
network.

Fig.6:
Distribution of the fitness value when the initial condition for $x_j$
is not fixed at -1, but is distributed over $[-1,1]$.  We choose the
evolved network as in Fig.5, and for each network we take 10000
initial conditions, and simulated the dynamics (1) without noise to
measure the fitness value $F$ after the system reached an attractor (as the
temporal average $400<t<500$).  The histogram is plotted with a bin size
0.1.

Fig.7:
Schematic representation of the basin structure, represented as a
process of climbing down a potential landscape.  $\Delta$ is the magnitude
of perturbation to jump over the barrier to a different attractor from the target.
Smooth landscape is evolved under high level noise (above), and
rugged landscape is evolved under low level noise (below).

Fig.8:
Schematic representation of an orbit in the phase space.
The solid curve is an original orbit from the initial condition (I)
to the target attractor (T).  Dashed curves are orbits perturbed by noise.
When orbits encounter turning points, they escape the original basin of attraction 
and may be caught in another attractor.  Mutations, on the other
hand, are able to move the position of turning points.

\pagebreak

\begin{figure}[tbp]
\begin{center}
\includegraphics[width=9.5cm,height=7.5cm]{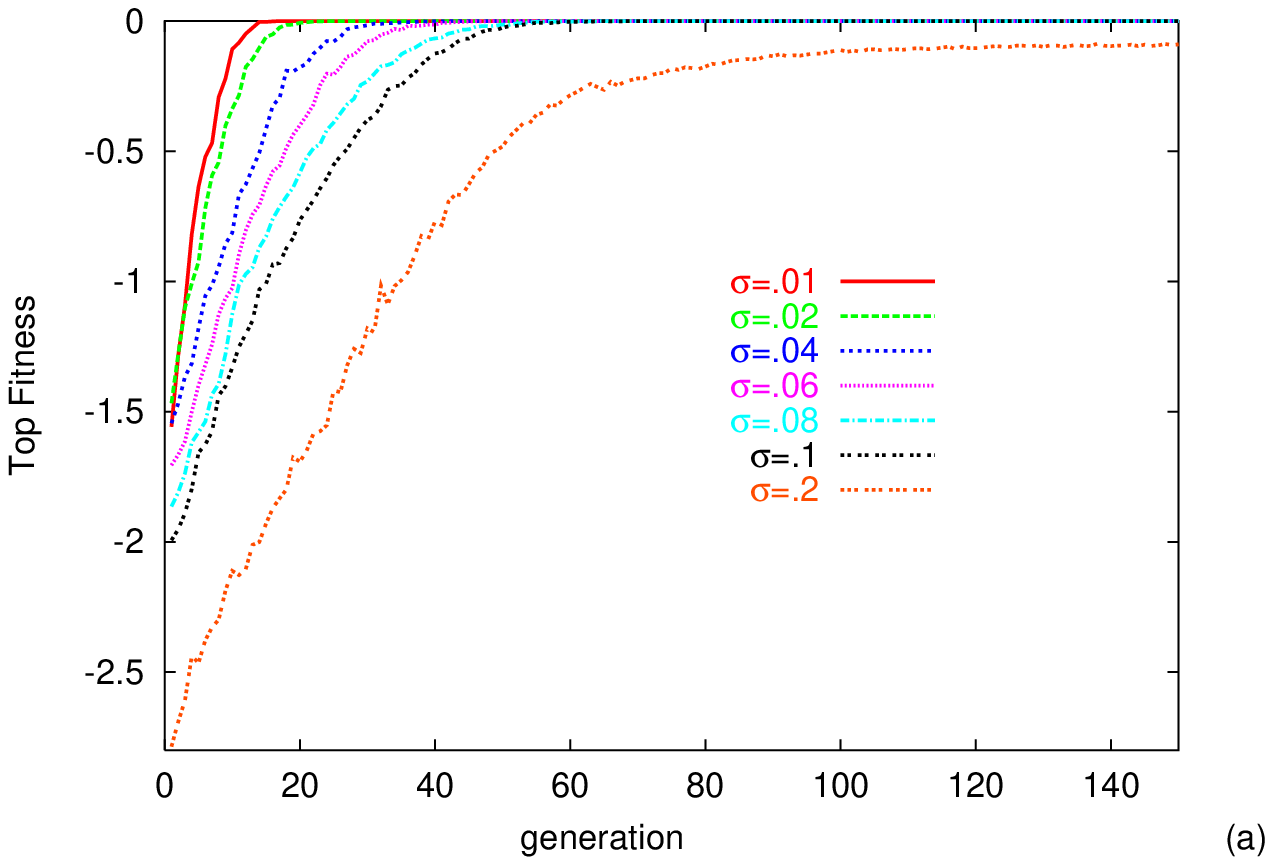}
\includegraphics[width=9.5cm,height=7.5cm]{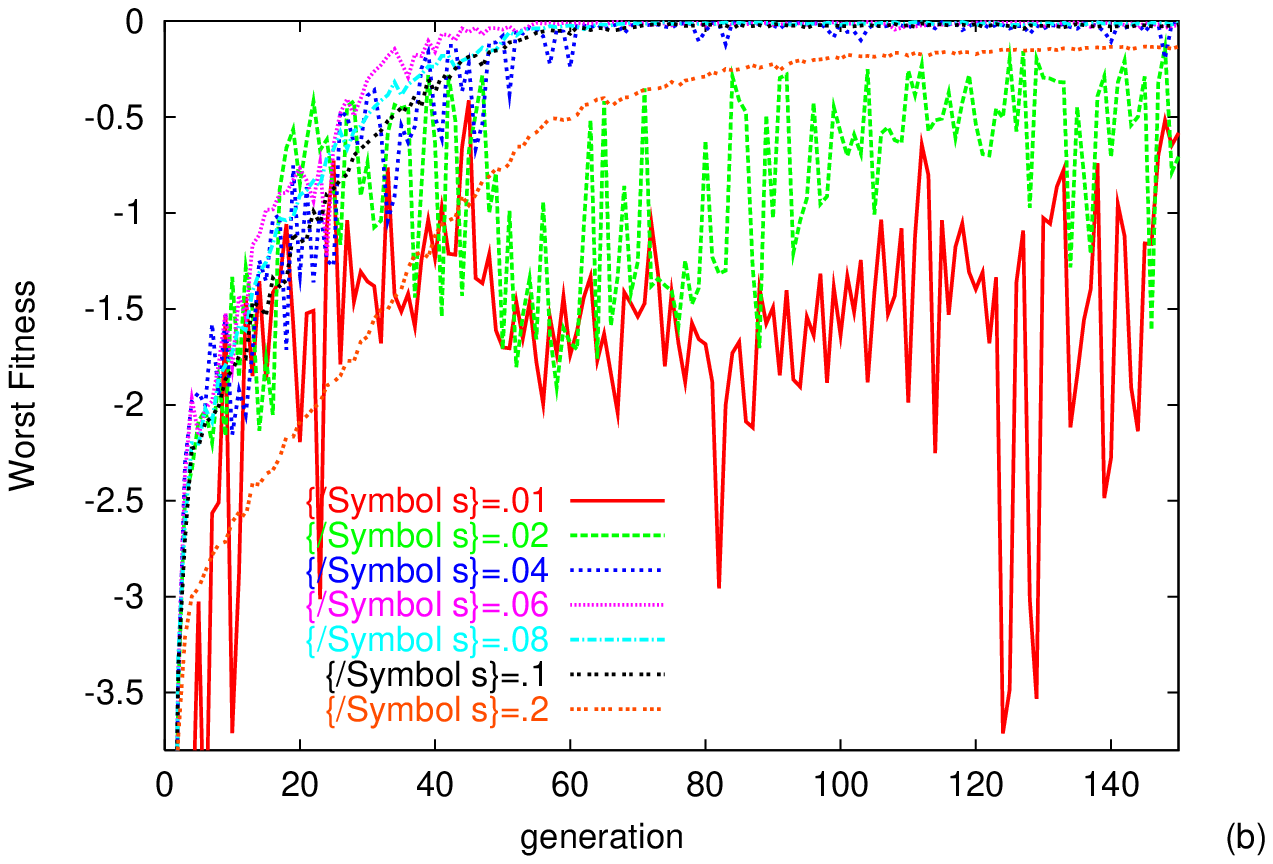}
\caption{
}
\end{center}
\end{figure}

\begin{figure}[tbp]
\begin{center}
\includegraphics[width=9.5cm,height=7.5cm]{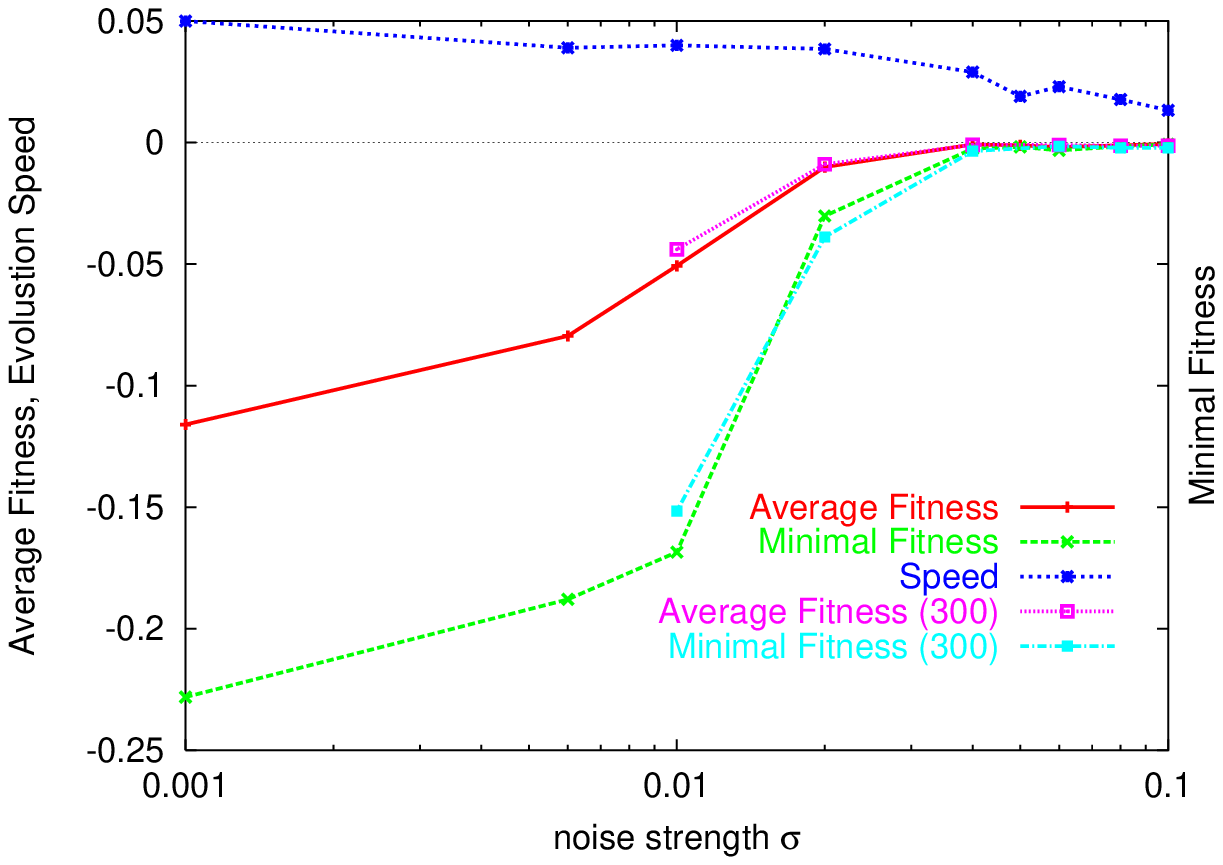}
\caption{
}
\end{center}
\end{figure}

\begin{figure}[tbp]
\begin{center}
\includegraphics[width=9.5cm,height=7.5cm]{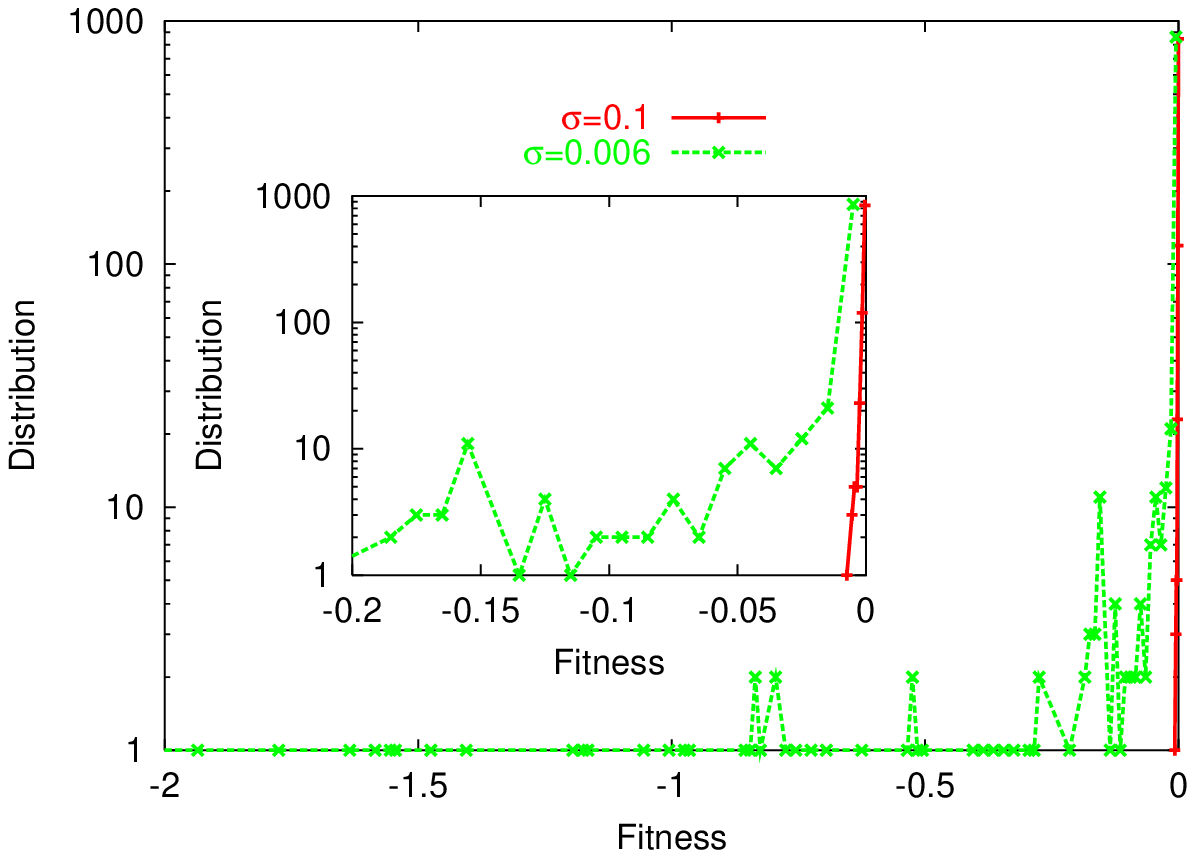}
\caption{
}
\end{center}
\end{figure}

\begin{figure}[tbp]
\begin{center}
\includegraphics[width=9.5cm,height=7.5cm]{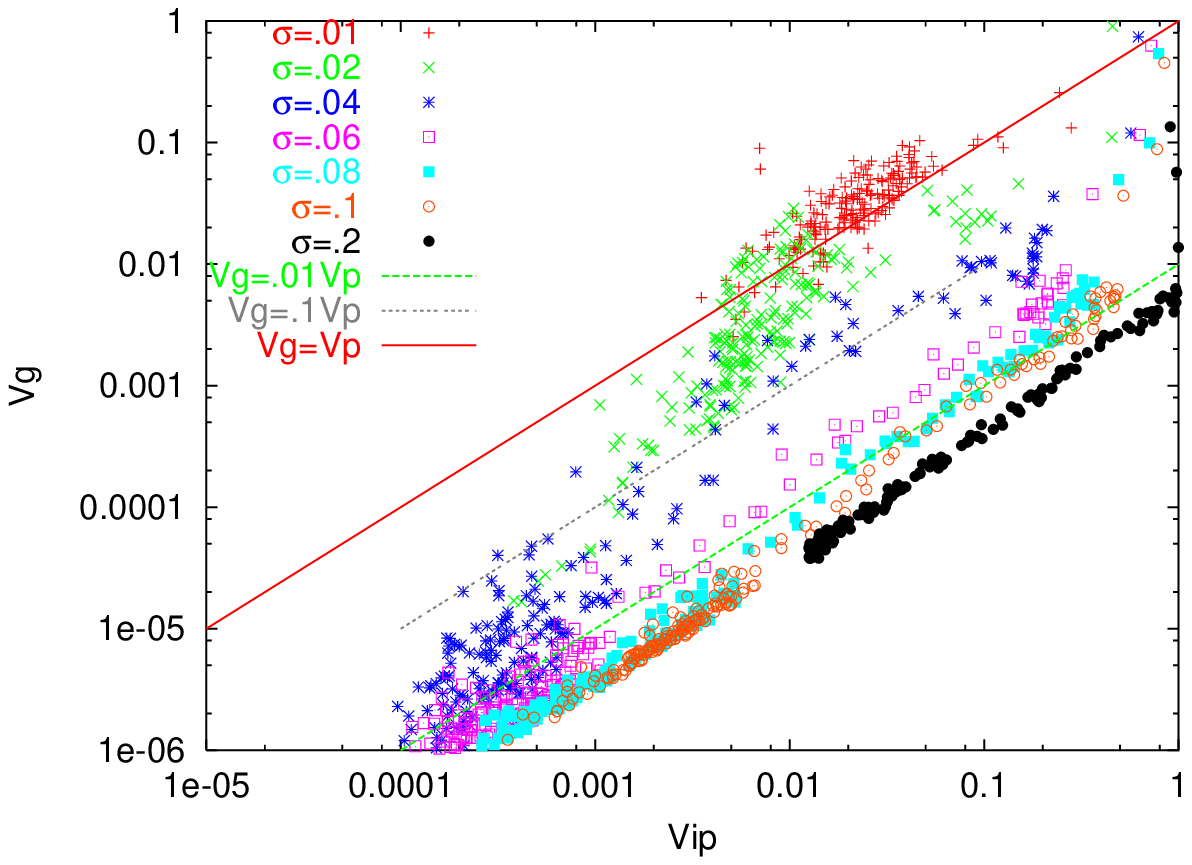}
\caption{
}
\end{center}
\end{figure}

\begin{figure}[tbp]
\begin{center}
\includegraphics[width=9.5cm,height=7.5cm]{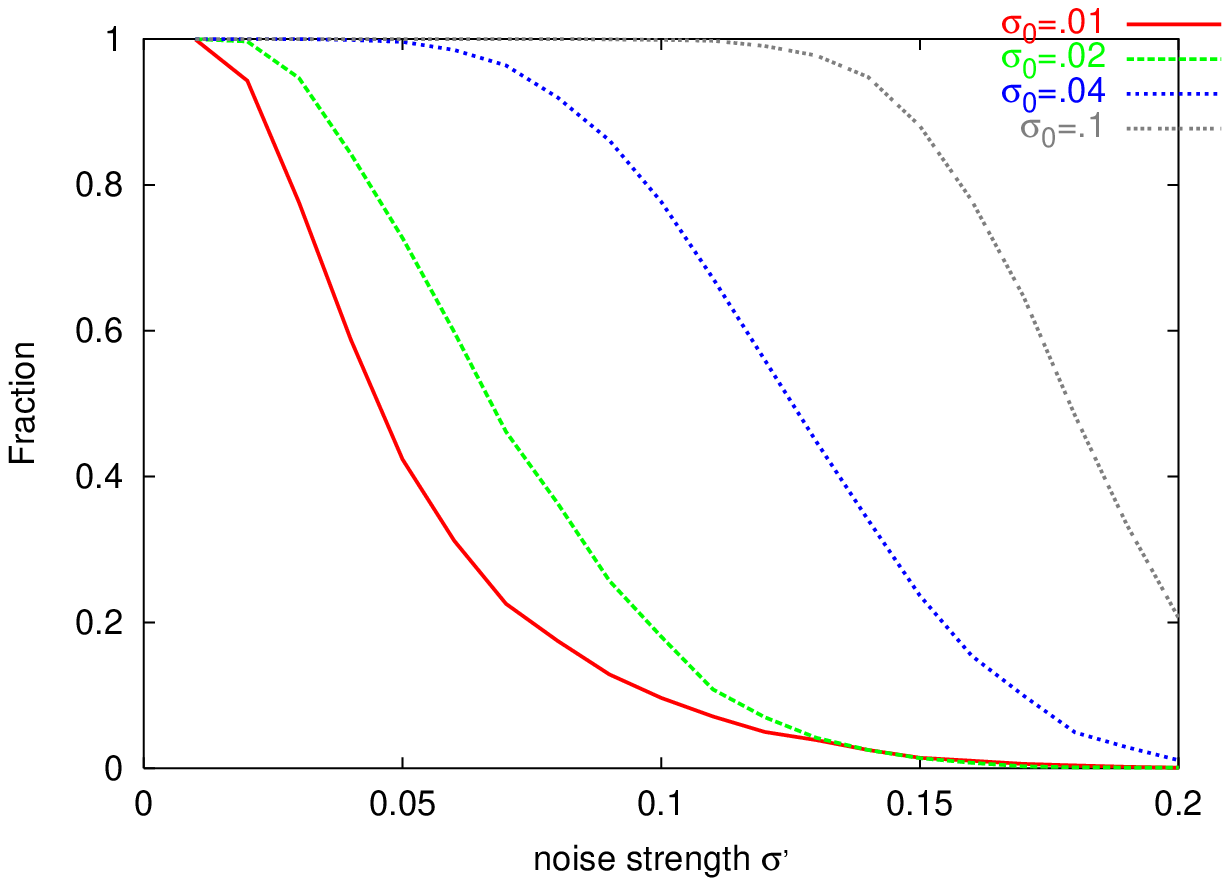}
\caption{ 
}
\end{center}
\end{figure}

\begin{figure}[tbp]
\begin{center}
\includegraphics[width=9.5cm,height=7.5cm]{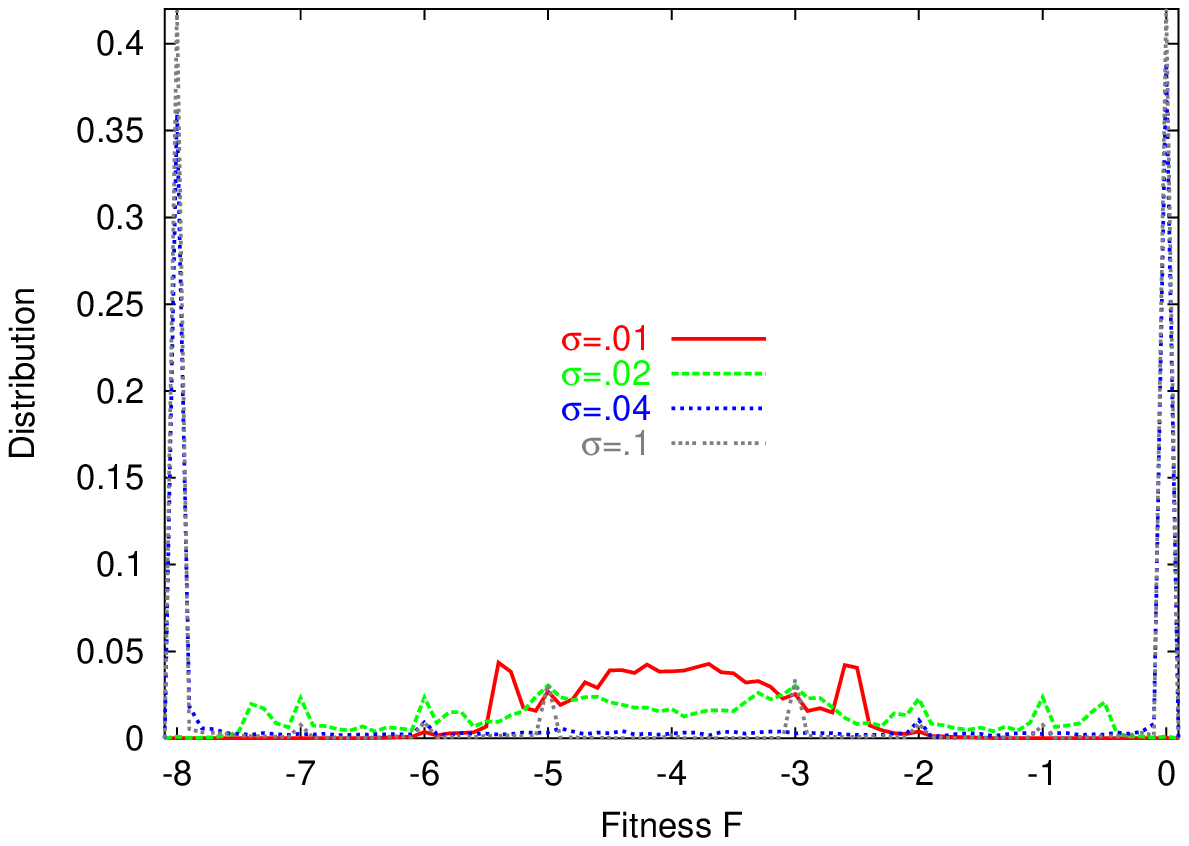}
\caption{ 
}
\end{center}
\end{figure}

\begin{figure}[tbp]
\begin{center}
\includegraphics[width=8.5cm,height=8.5cm]{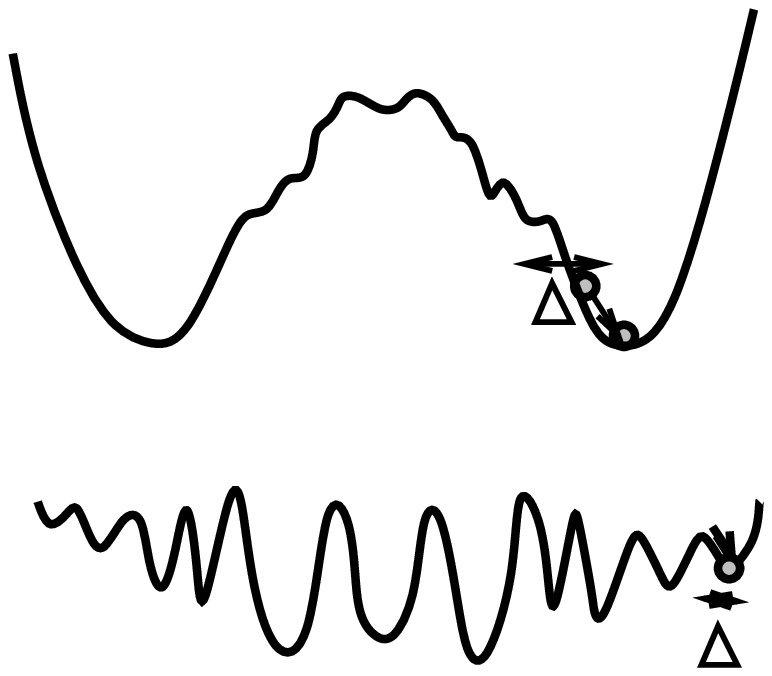}
\caption{ 
}
\end{center}
\end{figure}

\begin{figure}[tbp]
\begin{center}
\includegraphics[width=9.5cm,height=9.5cm]{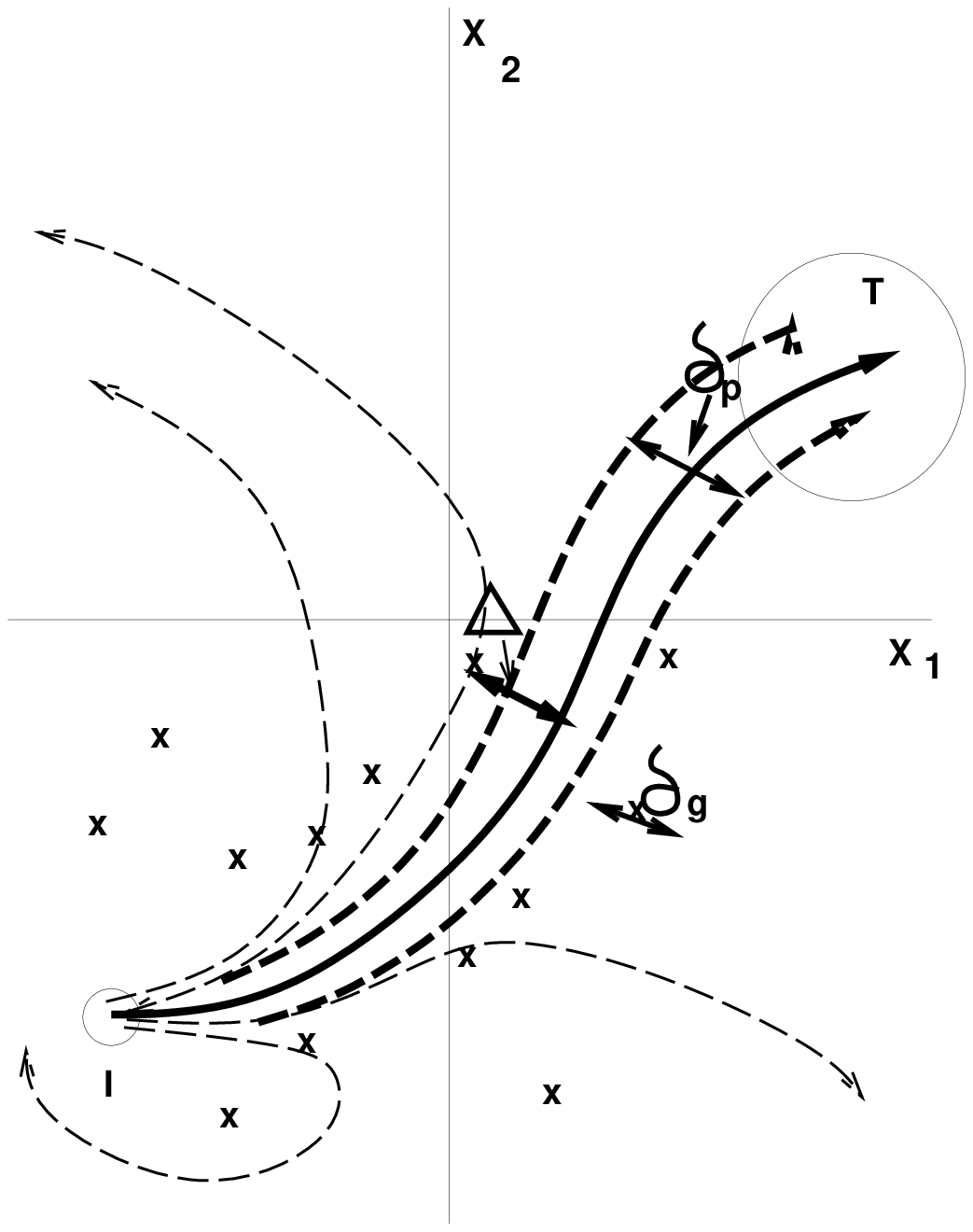}
\caption{ 
}
\end{center}
\end{figure}
\end{document}